\documentclass[11pt]{article}
\usepackage{graphicx}
\usepackage{xspace}
\usepackage{afterpage}

\pdfoutput=1

\newcommand{\ttbar}{\ensuremath{\mathrm{t\bar{t}}}\xspace}

\newcommand{\fUL}{fb$^{-1}$\xspace}

\newcommand{\pt}{\ensuremath{p_\mathrm{T}}\xspace}

\newcommand{\POWHEG}{\textsc{Powheg}\xspace}
\newcommand{\MCATNLO}{\textsc{MC@NLO}\xspace}
\newcommand{\AMCATNLO}{\textsc{aMC@NLO}\xspace}
\newcommand{\MADGRAPH}{\textsc{Madgraph}\xspace}
\newcommand{\PYTHIA}{\textsc{Pythia}\xspace}
\newcommand{\HERWIG}{\textsc{HERWIG}\xspace}
\newcommand{\SHERPA}{\textsc{Sherpa}\xspace}

\newcommand\pubnumber{}
\newcommand\pubdate{\today}

\def\institute{University of Rochester}

\def\Title#1{\begin{center} {\Large #1 } \end{center}}
\def\Author#1{\begin{center}{ \sc #1} \end{center}}
\def\Address#1{\begin{center}{ \it #1} \end{center}}

\newcommand\pubblock{\rightline{\begin{tabular}{l} \pubnumber\\
         \pubdate  \end{tabular}}}
\newenvironment{Abstract}{\begin{quotation}  }{\end{quotation}}
\newenvironment{Presented}{\begin{quotation} \begin{center} 
             PRESENTED AT\end{center}\bigskip 
      \begin{center}\begin{large}}{\end{large}\end{center} \end{quotation}}

\def\beq{\begin{equation}}
\def\eeq#1{\label{#1}\end{equation}}
\def\eeqn{\end{equation}}

\def\beqa{\begin{eqnarray}}
\def\eeqa#1{\label{#1}\end{eqnarray}}
\def\eeqan{\end{eqnarray}}

\let\bar=\overbar

\def\Dslash{\not{\hbox{\kern-4pt $D$}}}
\def\dslash{\not{\hbox{\kern-2pt $\del$}}}

\def\msb{{\bar{\ssstyle M \kern -1pt S}}}

\begin{document}

\begin{titlepage}
\pubblock

\vfill
\Title{Differential $t\bar{t}$ Cross Section Measurements as a Function of Variables other than Kinematics}
\vfill
\Author{Otto Hindrichs\\
For the ATLAS and CMS Collaborations}
\Address{\institute}
\vfill
\begin{Abstract}
An overview of cross section measurements as a function of jet multiplicities and jet kinematics in association with \ttbar production is presented. Both the ATLAS and the CMS collaborations performed a large number of measurements at different center-of-mass energies of the LHC using various \ttbar decay channels. Theoretical predictions of these quantities usually rely on parton shower simulations that strongly depends on tunable parameters and come with large uncertainties. The measurements are compared to various theoretical descriptions based on different combinations of matrix-element calculations and parton-shower models.         
\end{Abstract}
\vfill
\begin{Presented}
$9^{th}$ International Workshop on Top Quark Physics\\
Olomouc, Czech Republic,  September 19--23, 2016
\end{Presented}
\vfill
\end{titlepage}
\def\thefootnote{\fnsymbol{footnote}}
\setcounter{footnote}{0}

\section{Introduction}

At the LHC a large number of top quark-antiquark pairs (\ttbar) are produced together with additional jets. A measurement of jet multiplicities and jet kinematics provides insights in standard model QCD. However, while on the one hand such measurements have been performed with high precision by the ATLAS\cite{AT} and CMS\cite{CMS} collaborations, one the other hand the theoretical prediction is difficult. Full fixed order calculations are challenging for high jet multiplicities and have to be matched to parton-shower simulations. We will show that the tuning of the parton shower and the matching between matrix-element calculation and parton shower has large impact on the theoretical description.

The measured results are usually shown for a limited range of the phase space, which is similar to the experimental acceptance, and are presented at particle level. The particle level is described by long-living particles, which directly affect the measurements within the detectors. Such a definition of the measured observables minimizes the dependency on theoretical extrapolations.   

\section{Measurements}

\begin{figure}[!htb]
\centering
\includegraphics[width=0.3\textwidth]{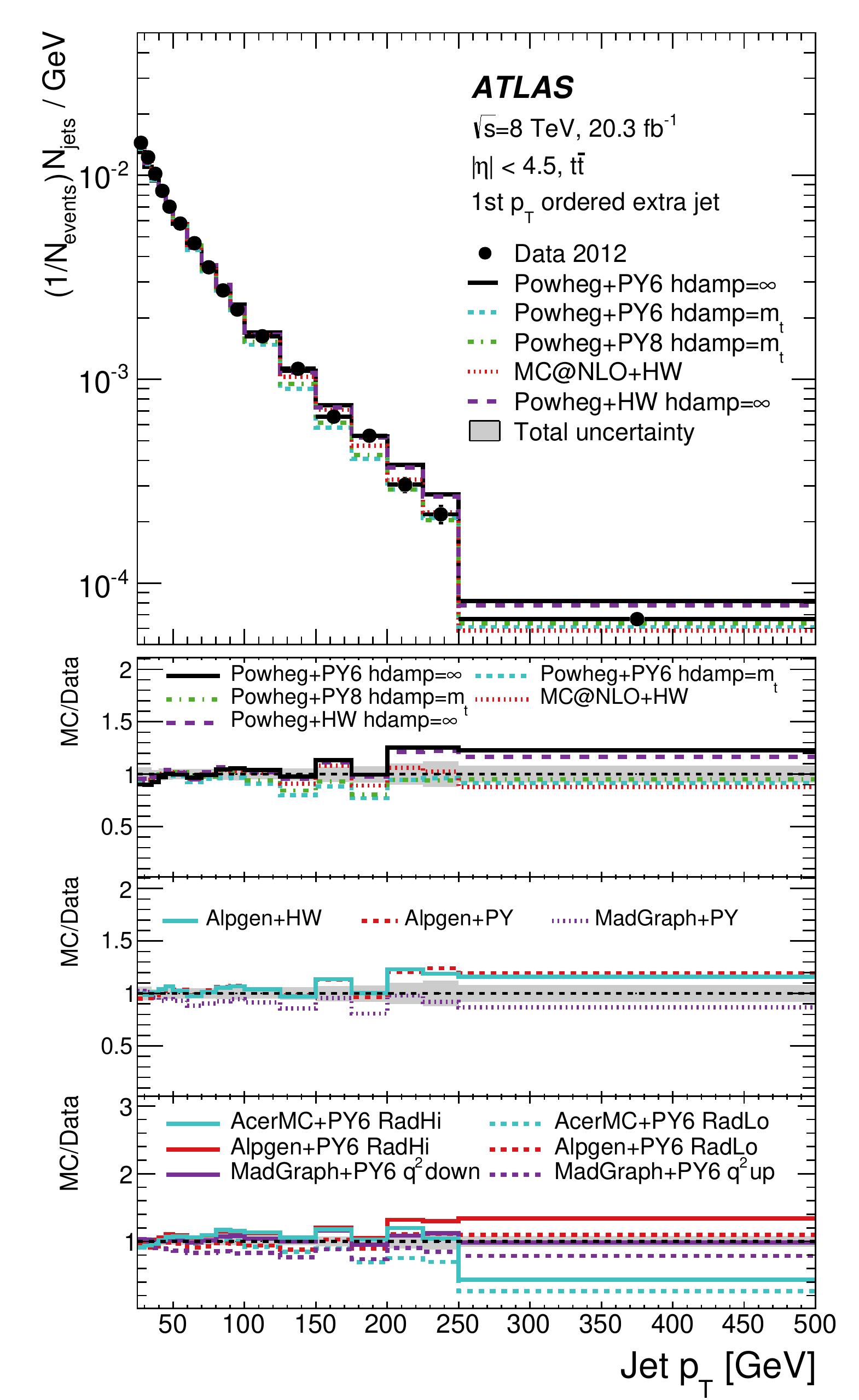}
\includegraphics[width=0.3\textwidth]{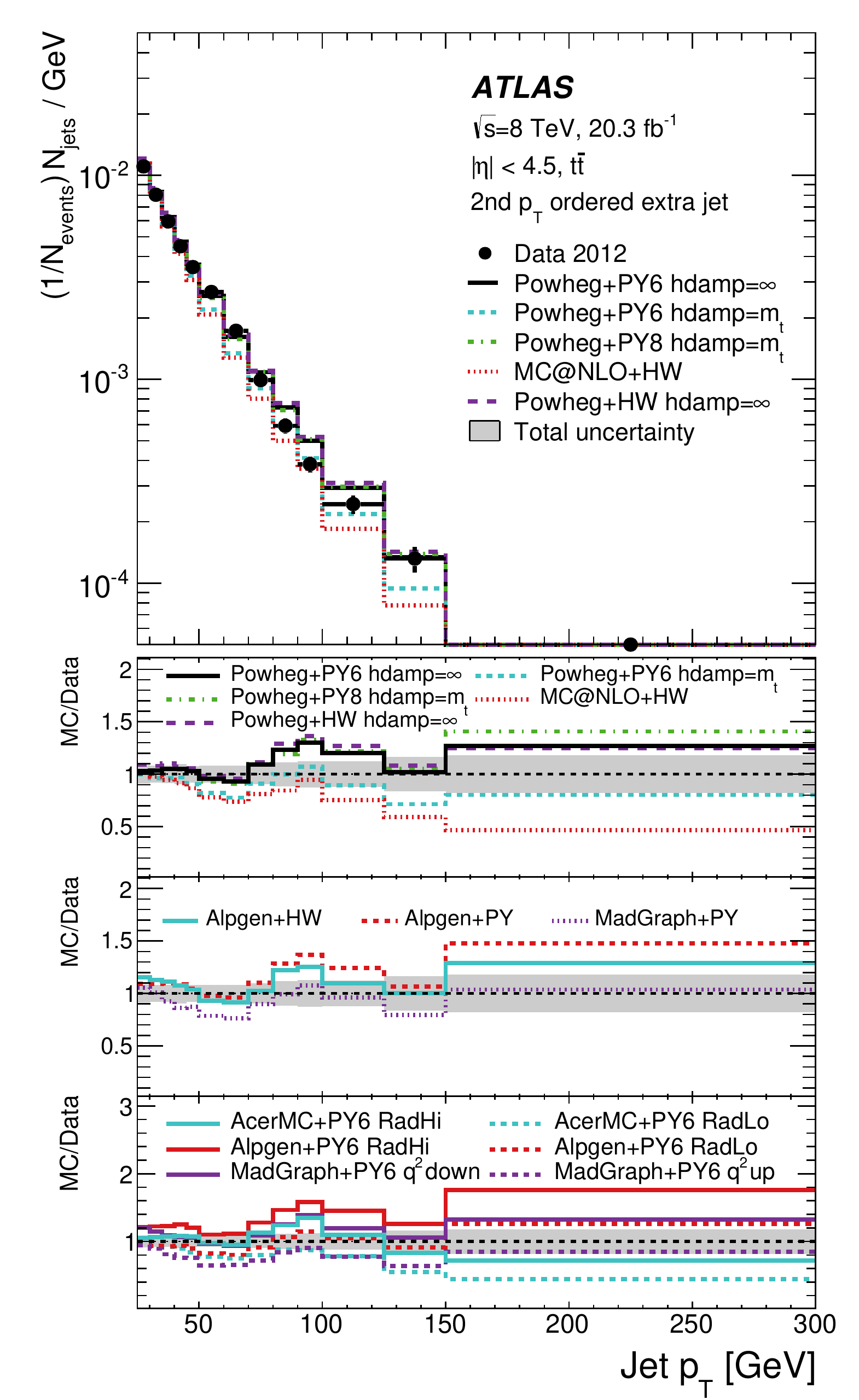}
\includegraphics[width=0.3\textwidth]{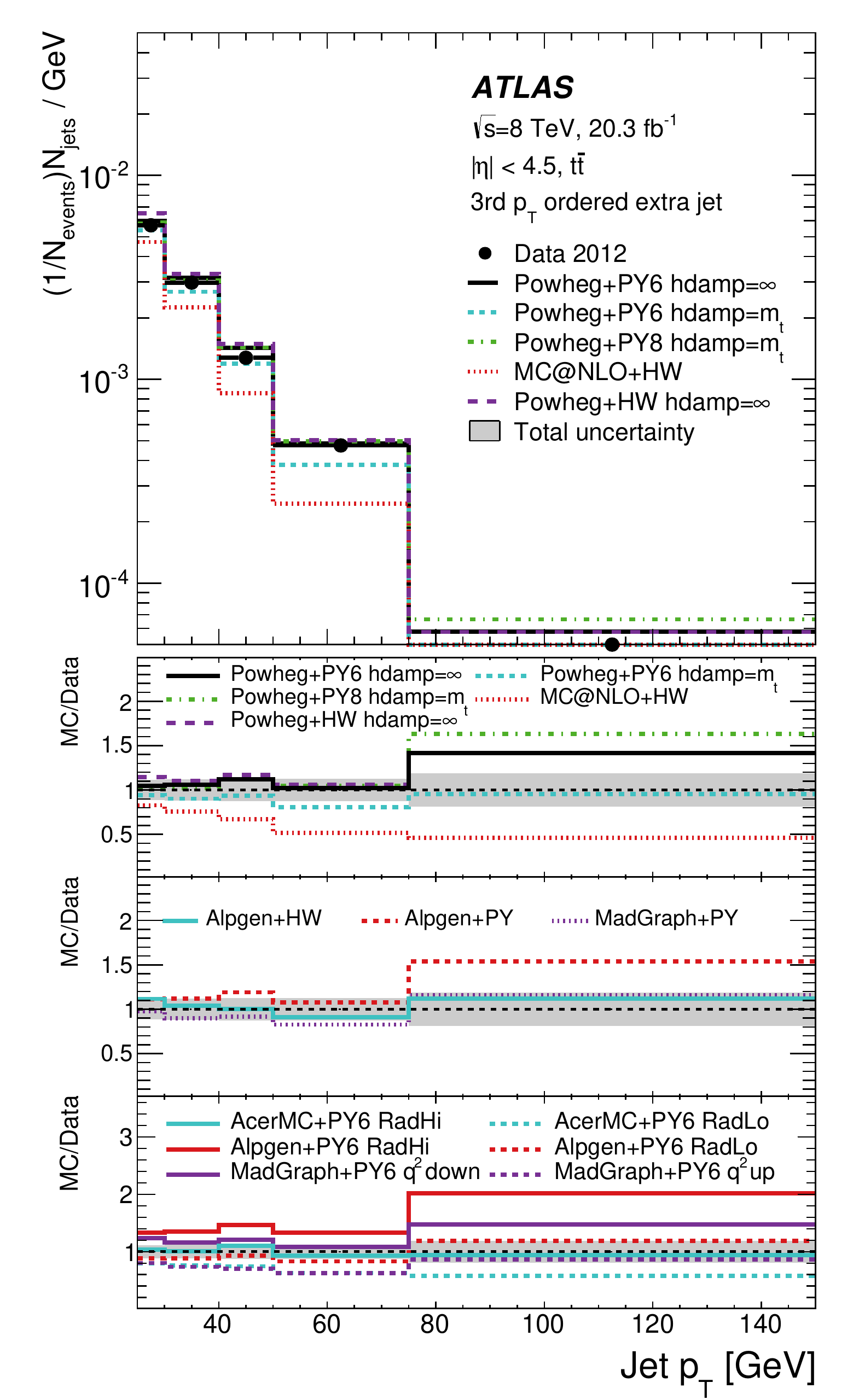}
\caption{The top panels show the normalized differential cross section as a function of the \pt of additional jets ranked by their \pt. The measured results are compared to various theoretical predictions. The lower panels show ratios of various theoretical predictions to the measured result\,\cite{AT1}.} 
\label{FAT1a}
\end{figure}

Based on an integrated luminosity of 20.3\,\fUL recorded at 8\,TeV ATLAS\,\cite{AT1} measured the jet multiplicity and the transverse momentum \pt of the additional jets ordered by their rank where the additional jet with the highest \pt has rank one, with the second highest \pt rank two, and so on. In this measurement \ttbar events in the dilepton channel are selected with one electron and one muon in the final state. In addition, at least two jets identified as b jets are required. The two b jets most compatible with the b jet assumption are assigned to the \ttbar system. All other jets with $\pt > 25$\,GeV and a pseudorapidity $|\eta| < 4.5$ are considered as additional jets. The measured distributions are unfolded simultaneously in the \pt and the rank of the jets. The results of the differential cross section as a function of \pt for the three highest ranked jets are shown in Fig.\,\ref{FAT1a}. \POWHEG and \MCATNLO perform a NLO calculation of \ttbar production. The results are further combined with the parton shower simulations of \PYTHIA and \HERWIG, respectively, i.e., the radiation of one additional jet is described with LO precision while all other additional jets are based on the parton-shower simulations. A comparison to the measurement shows that the leading jet is well described and only marginally affected by the variation of $h_\mathrm{damp}$ in the \POWHEG simulations.     

\begin{figure}[!htb]
\centering
\includegraphics[width=0.32\textwidth]{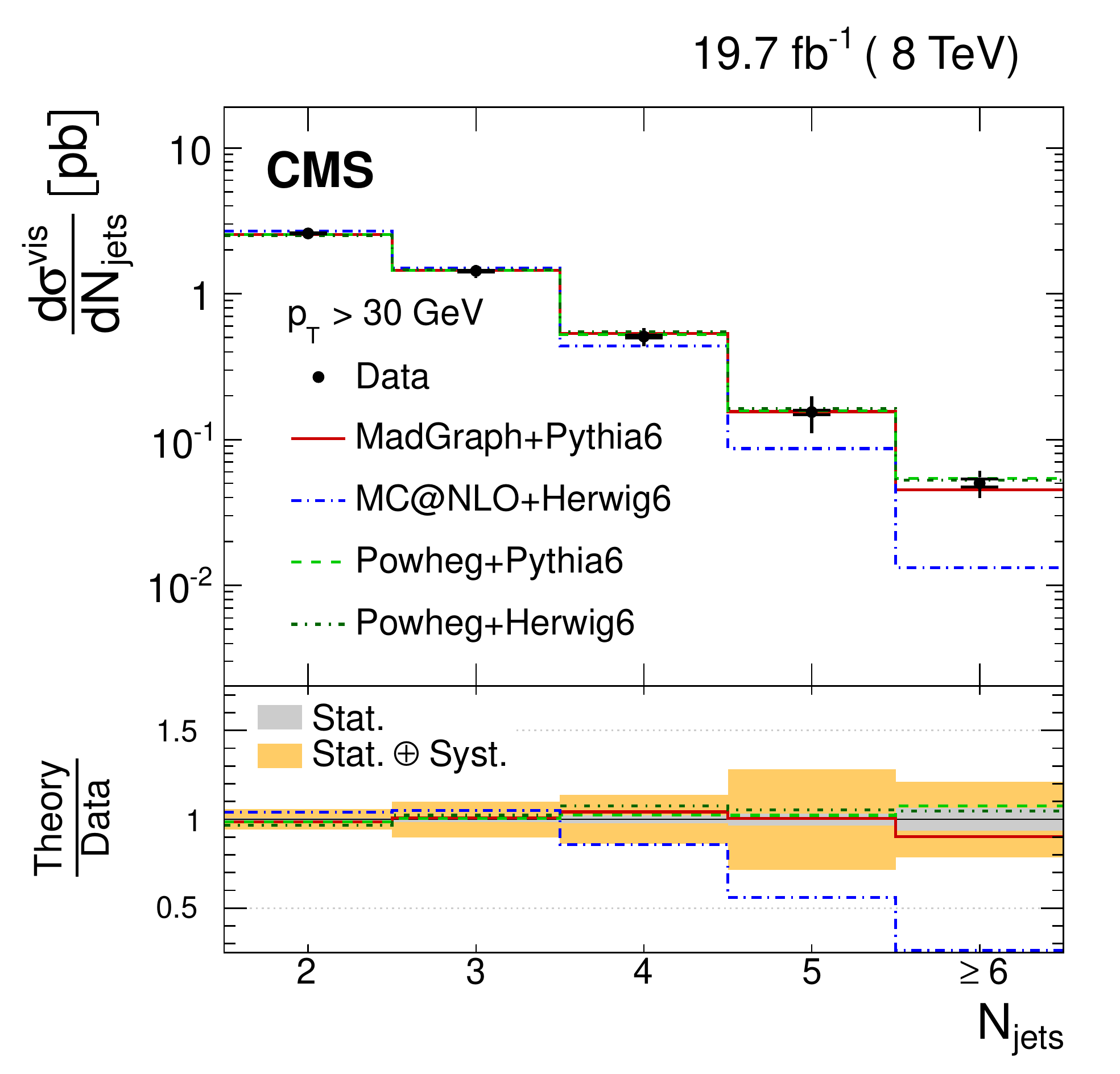}
\includegraphics[width=0.32\textwidth]{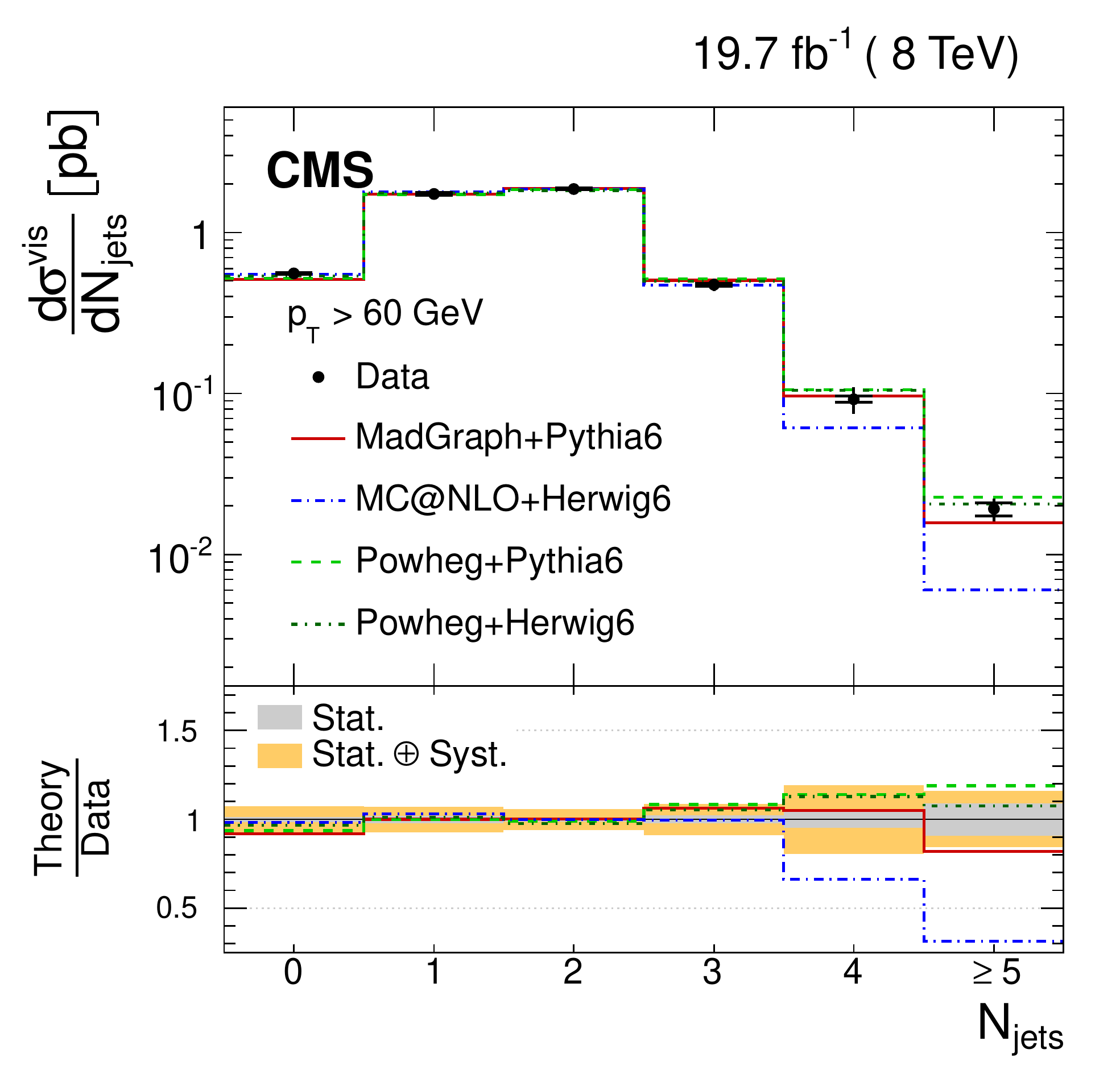}
\includegraphics[width=0.32\textwidth]{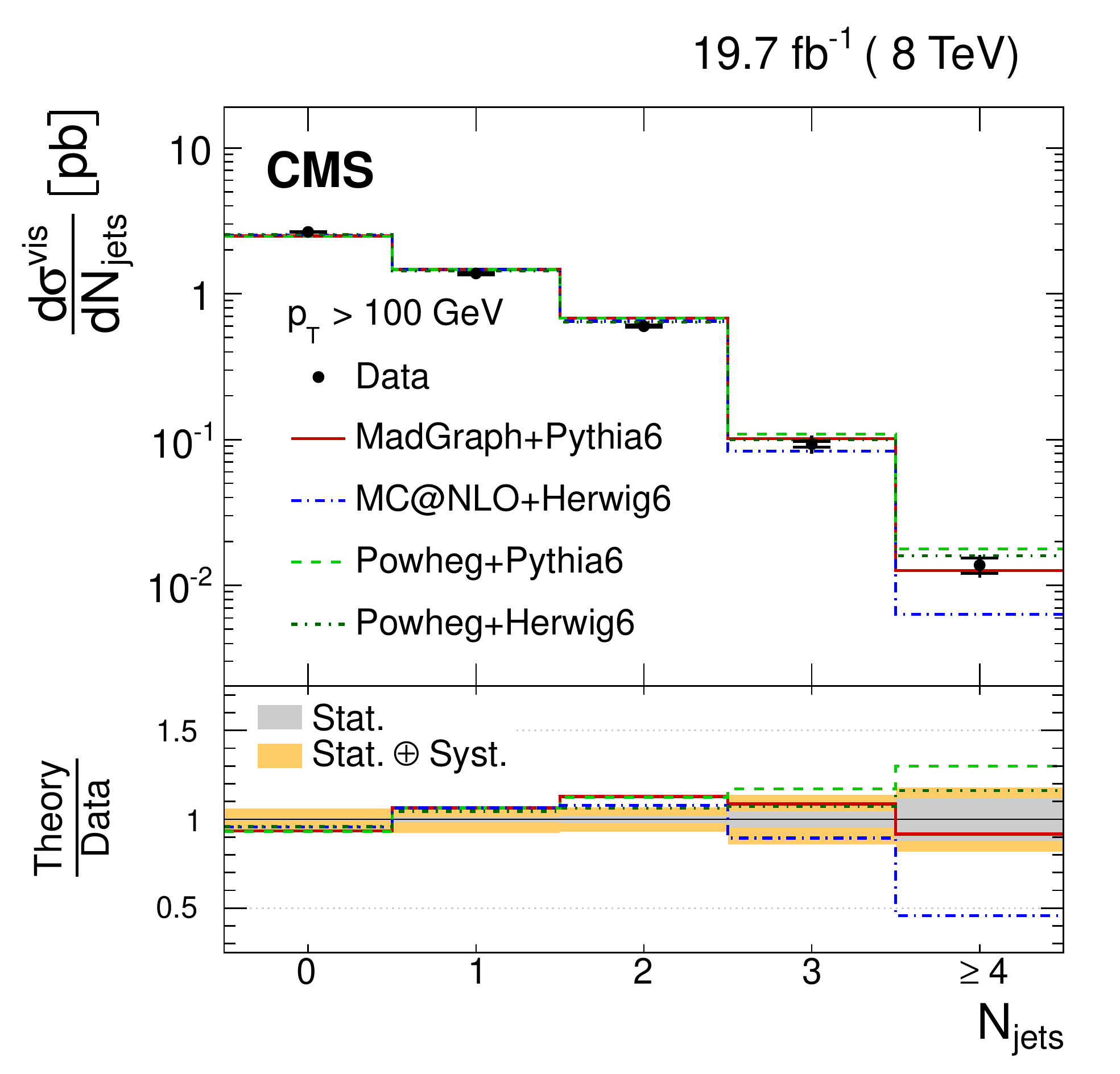}
\caption{Measured multiplicities for jets with \pt above 30\,GeV(left), 60\,GeV(middle), and 100\,GeV(right) compared to various theoretical predictions. The lower panels show the corresponding ratios of the predictions to the measured result\,\cite{CMS1}.} 
\label{FCMS1a}
\end{figure}

A similar analysis is performed by CMS\,\cite{CMS1}. Based on an integrated luminosity of 19.7\,\fUL recorded at 8\,TeV. In this measurement the ee, $\mu\mu$, and $e\mu$ channels are used. At least two jets including at least one b-tagged jet with $\pt > 20$ and $|\eta| < 2.4$ are required. Afterwards a kinematic reconstruction of the \ttbar system is performed to identify the b jets from top quark decays. The unfolded number of jets above various \pt thresholds are shown in Fig.\,\ref{FCMS1a}. The measurement is well described by the \POWHEG and the \MADGRAPH simulations where \MADGRAPH is used to calculate the processes of \ttbar plus up to three additional partons with LO accuracy. These are merged and combined with the parton-shower simulation. The combination of \MCATNLO and \HERWIG underestimates the jet multiplicities; an observation consistent with the  ATLAS result.     

\begin{figure}[!htb]
\centering
\includegraphics[width=0.4\textwidth]{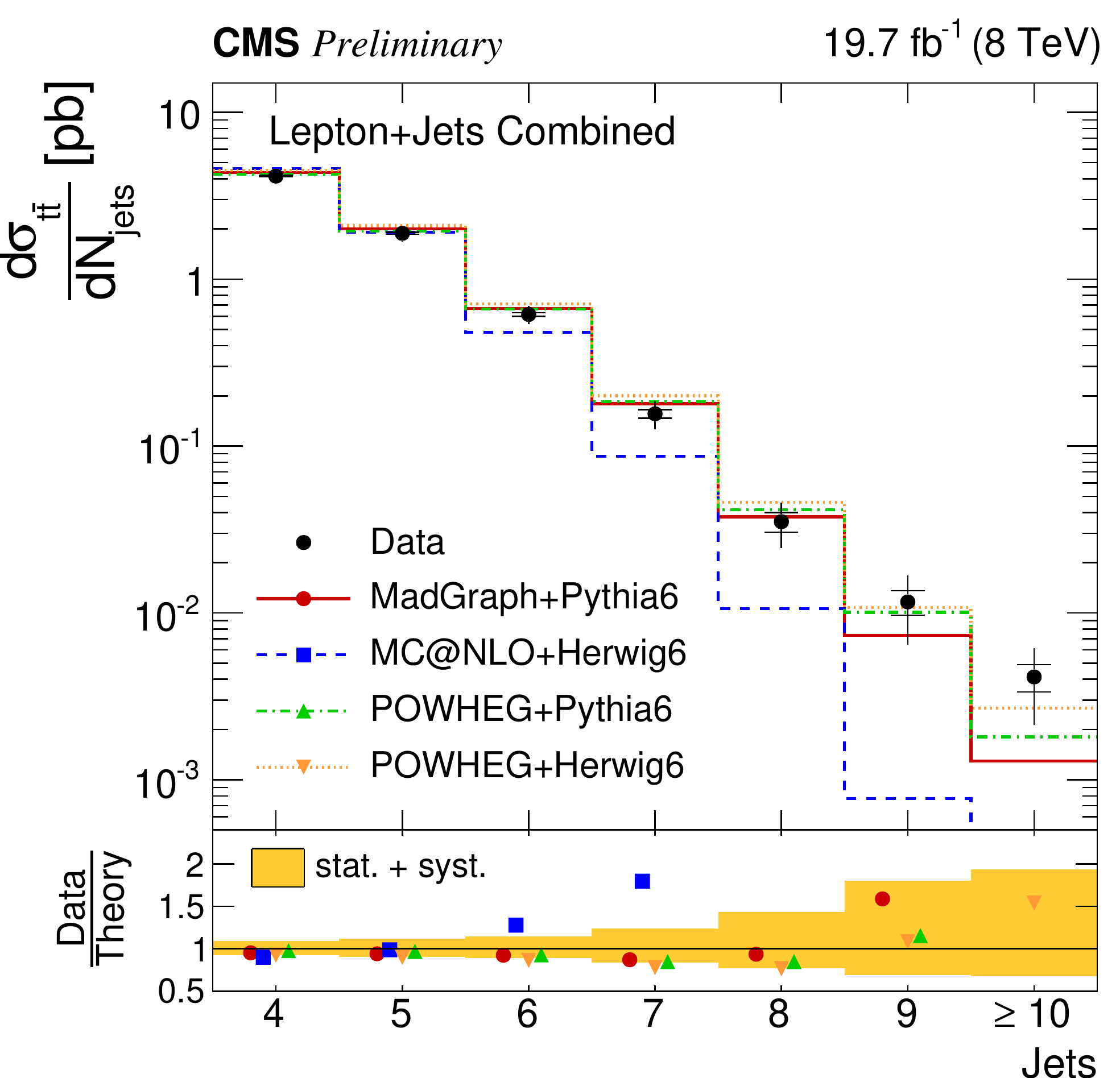}
\caption{Multiplicities for jets with $\pt > 30$\,GeV measured in the e/$\mu$ + jets channel compared to various theoretical predictions. The lower panel shows the corresponding ratios of the measured result to the predictions\,\cite{CMS2}.} 
\label{FCMS2a}
\end{figure}

The jet multiplicities are also measured in the e/$\mu$ + jets channel by CMS\,\cite{CMS2}. Events with a single isolated lepton together with at least four jets with $\pt > 30$\,GeV are selected. At least two of the jets have to be identified as b jets. The jet multiplicity measured in this channel is shown in Fig.\,\ref{FCMS2a}. The agreements between measurement and simulations are consistent with those in the dilepton channel.

\begin{figure}[!htb]
\centering
\includegraphics[width=0.32\textwidth]{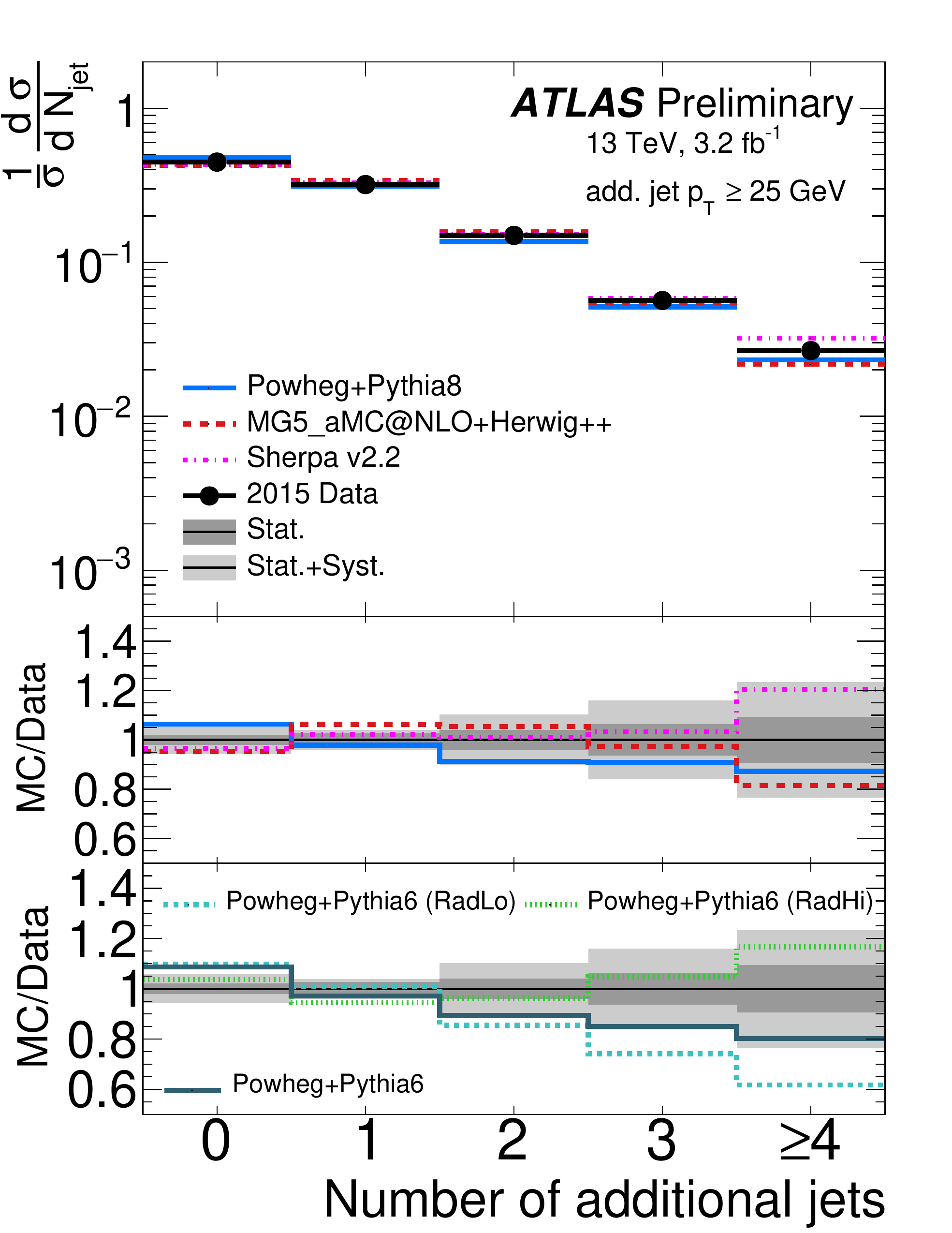}
\includegraphics[width=0.32\textwidth]{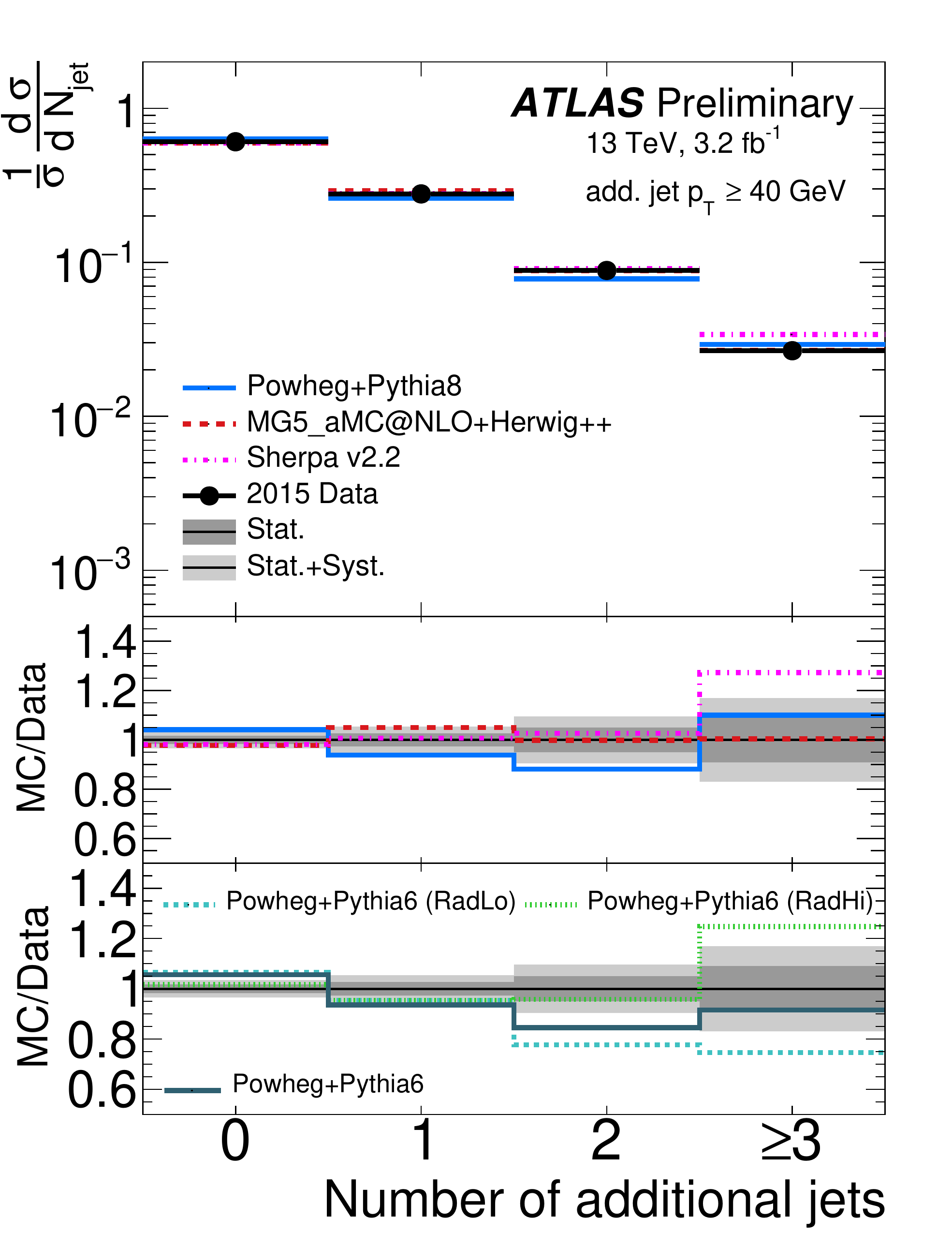}
\includegraphics[width=0.32\textwidth]{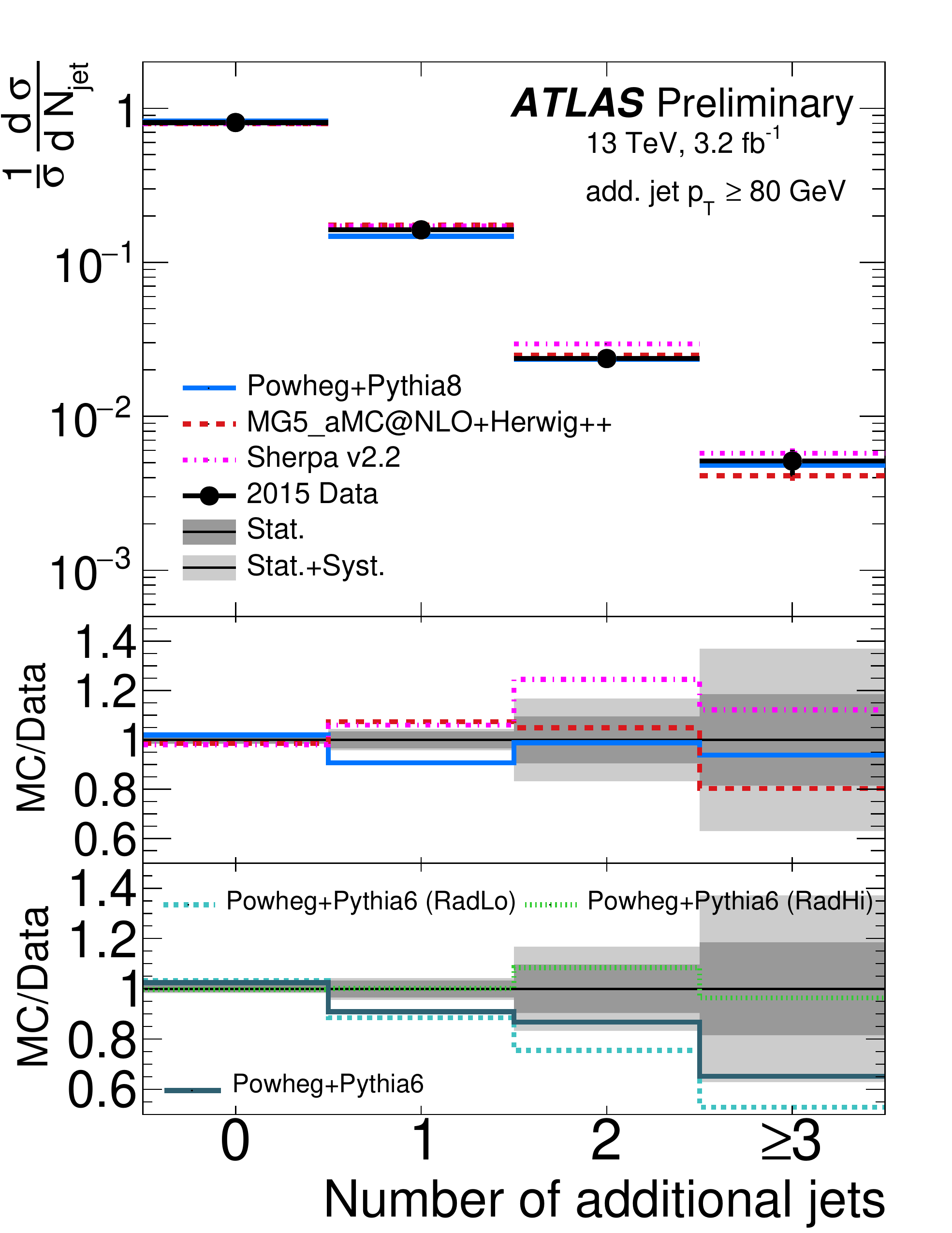}
\caption{Normalized distributions of multiplicities for jets with \pt above 25\,GeV(left), 40\,GeV(middle), and 80\,GeV(right) compared to various theoretical predictions are shown in the top panels. The middle and bottom panels show the ratios of different predictions to the measured result\,\cite{AT2}.} 
\label{FAT2a}
\end{figure}

Meanwhile first results measured at 13\,TeV center-of-mass energy are available. Based on 3.7\,\fUL ATLAS\,\cite{AT2} performed a measurement of jet multiplicities in the dilepton channel selecting e$\mu$ events with two b-tagged jets. The result is shown in Fig.\,\ref{FAT2a}. The measurement is well described by a \POWHEG, a \AMCATNLO, and a \SHERPA simulation. The effect of the parton shower tuning and matching is demonstrated by varying the parameters within their uncertainties to enhance (RadHi) or reduce (RadLo) the radiation of additional jets. This demonstrates the sensitivity of the calculations to the tuning parameters such as the choices of scales for the parton-shower calculation and the matching.

\begin{figure}[tb]
\centering
\includegraphics[width=0.9\textwidth]{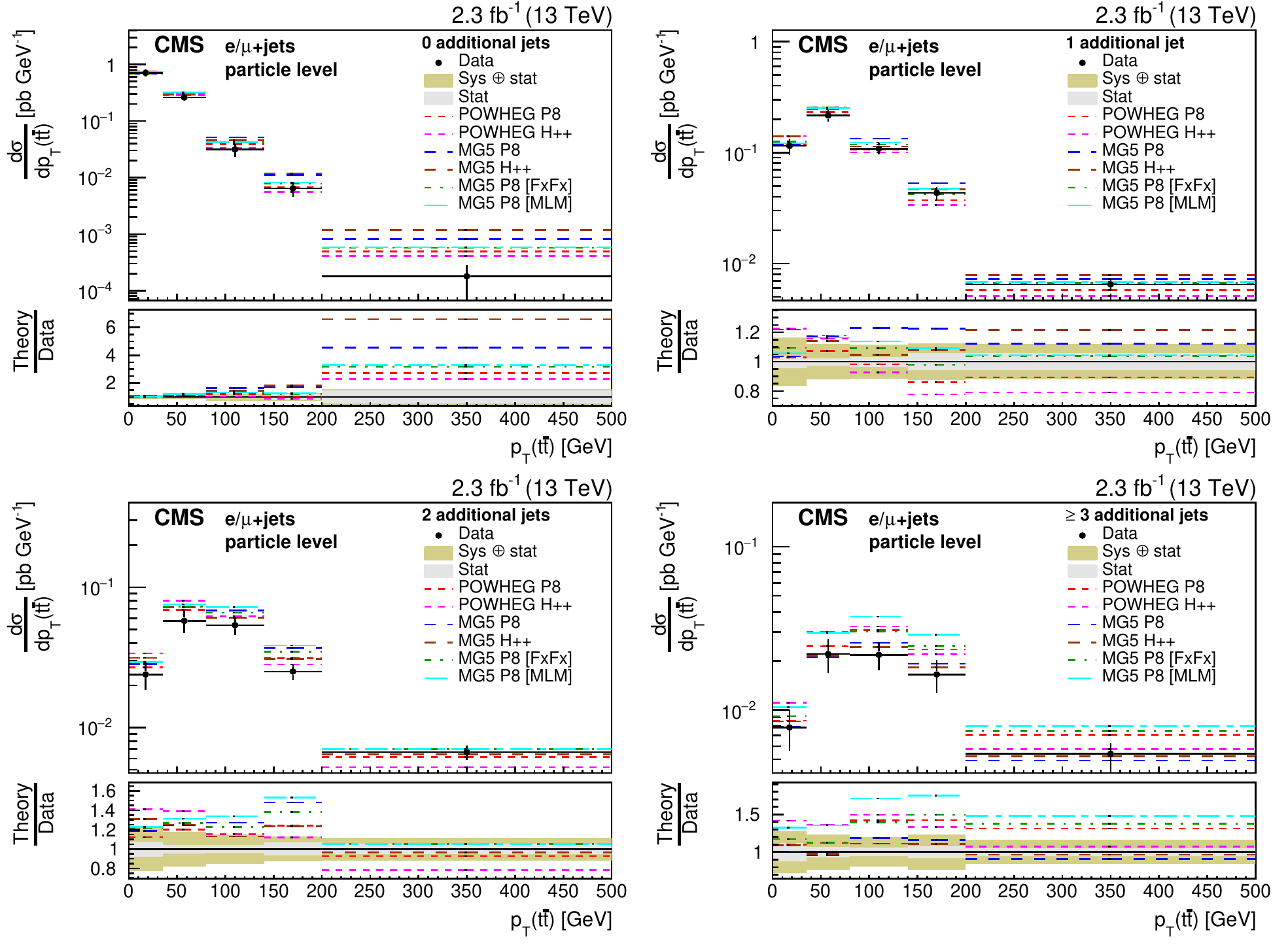}
\caption{Measured differential cross section as a function of the \pt of the \ttbar system for different numbers of additional jets compared to various theoretical predictions. The lower panels show the corresponding ratios of the predictions to the measured results\,\cite{CMS3}.} 
\label{FCMS3a}
\end{figure}

Finally, CMS\,\cite{CMS3} performed a measurement of differential cross sections as a function of kinematic variables of the \ttbar system for various jet multiplicities. This analysis is based on 2.3\,\fUL at 13\,TeV in the $e/\mu$ + jets channel. In addition to the lepton at least four jets are required out of which at least two are identified as b jets. Based on the constraints of top quark and W boson masses particle-level top quarks are defined. These are reconstructed in the data and unfolded simultaneously in \pt of the \ttbar system and jet multiplicity. The result is shown in Fig.\,\ref{FCMS3a}. Naively none of the simulations seems to describes the data well, but taking into account the large uncertainties in the predictions, as discussed in the last paragraph, one would not expect a better agreement. 

For certain precision measurements and searches beyond the standard model it is helpful to select a simulation that describes the data as well as possible keeping in mind the uncertainties and the risk of over-tuning, which could compromise the simulation towards nonstandard model physics.

\end{document}